\overfullrule=0pt
\input harvmac
\input amssym.tex
\def\a{{\alpha}}

\def\ab{{\overline\alpha}}

\def\l{{\lambda}}

\def\lh{{\widehat\lambda}}

\def\b{{\beta}}
\def\bb{{\overline\beta}}

\def\g{{\gamma}}
\def\gb{{\overline\gamma}}

\def\d{{\delta}}
\def\db{{\overline\delta}}

\def\r{{\rho}}
\def\rb{{\overline\rho}}

\def\N{{\nabla}}
\def\Nb{{\overline\nabla}}

\def\O{{\Omega}}

\def\Oh{{\widehat\O}}
\def\o{{\omega}}

\def\oh{{\widehat\omega}}

\def\p{{\partial}}
\def\pb{{\overline\partial}}
\def\t{{\theta}}

\def\oh{{\widehat\o}}
\def\L{{\Lambda}}

\def\Pib{{\overline\Pi}}

\def\Gb{\overline\Gamma}
\def\S{{\Sigma}}
\def\St{{\widetilde\Sigma}}

\def\Th{{\widehat T}}

\def\NN{{\overline N}}

\def\zz{{\overline z}}

\def\yy{{\overline y}}

\baselineskip 12pt

\Title{ \vbox{\baselineskip 12pt
}}
{\vbox{\centerline
{The non-minimal type II pure spinor string}
\bigskip
\centerline{in a curved background}
}}
\smallskip
\centerline{Osvaldo Chandia\foot{e-mail: ochandiaq@gmail.com}, }
\bigskip
\centerline{ \it Departamento de Ciencias, Facultad de Artes Liberales, Universidad Adolfo Ib\'a\~nez}
\centerline{\it Diagonal Las Torres 2640, Pe\~nalol\'en, Santiago, Chile} 

\bigskip
\bigskip
\bigskip
\bigskip

\noindent
The pure spinor superstring in a type II curved background is considered. In order to define reparametrization ghosts, non-minimal pure spinor variables have to be present in the formalism. The BRST transformations of the non-minimal variables are obtained. It is found that the BRST transformations of a set of world-sheet variables have the form of the corresponding transformations in flat spacetime, up to a field-dependent Lorentz rotation. This simplify the construction of the reparametrization $b$ ghosts.

\Date{October 2019}


\lref\BerkovitsFE{
  N.~Berkovits,
  ``Super Poincare covariant quantization of the superstring,''
JHEP {\bf 0004}, 018 (2000).
[hep-th/0001035].
}

\lref\BerkovitsNN{
  N.~Berkovits,
  ``Cohomology in the pure spinor formalism for the superstring,''
JHEP {\bf 0009}, 046 (2000).
[hep-th/0006003].
}

\lref\BerkovitsMX{
  N.~Berkovits and O.~Chandia,
  ``Lorentz invariance of the pure spinor BRST cohomology for the superstring,''
Phys.\ Lett.\ B {\bf 514}, 394 (2001).
[hep-th/0105149].
}

\lref\BerkovitsPH{
  N.~Berkovits and B.~C.~Vallilo,
  ``Consistency of superPoincare covariant superstring tree amplitudes,''
JHEP {\bf 0007}, 015 (2000).
[hep-th/0004171].
}

\lref\MafraWQ{
  C.~R.~Mafra,
  ``Superstring Scattering Amplitudes with the Pure Spinor Formalism,''
[arXiv:0902.1552 [hep-th]].
}

\lref\BerkovitsQX{
  N.~Berkovits and O.~Chandia,
  ``Massive superstring vertex operator in D = 10 superspace,''
JHEP {\bf 0208}, 040 (2002).
[hep-th/0204121].
}

\lref\BerkovitsYR{
  N.~Berkovits and O.~Chandia,
  ``Superstring vertex operators in an $AdS_5\times S^5$ background,''
Nucl.\ Phys.\ B {\bf 596}, 185 (2001).
[hep-th/0009168].
}

\lref\BerkovitsBT{
  N.~Berkovits,
  ``Pure spinor formalism as an N=2 topological string,''
JHEP {\bf 0510}, 089 (2005).
[hep-th/0509120].
}

\lref\BerkovitsUE{
  N.~Berkovits and P.~S.~Howe,
  ``Ten-dimensional supergravity constraints from the pure spinor formalism for the superstring,''
Nucl.\ Phys.\ B {\bf 635}, 75 (2002).
[hep-th/0112160].
}

\lref\ChandiaIMA{
  O.~Chandia,
  ``The Non-minimal Heterotic Pure Spinor String in a Curved Background,''
JHEP {\bf 1403}, 095 (2014).
[arXiv:1311.7012 [hep-th]].
}

\lref\BerkovitsAMA{
  N.~Berkovits and O.~Chandia,
  ``Simplified Pure Spinor $b$ Ghost in a Curved Heterotic Superstring Background,''
JHEP {\bf 1406}, 001 (2014).
[arXiv:1403.2429 [hep-th]].
}

\lref\BerkovitsPLA{
  N.~Berkovits,
  ``Dynamical twisting and the b ghost in the pure spinor formalism,''
JHEP {\bf 1306}, 091 (2013).
[arXiv:1305.0693 [hep-th]].
}

\lref\BedoyaIC{
  O.~A.~Bedoya and O.~Chandia,
  ``One-loop Conformal Invariance of the Type II Pure Spinor Superstring in a Curved Background,''
JHEP {\bf 0701}, 042 (2007). 
[hep-th/0609161].
}

\lref\ChandiaPAJ{
  O.~Chandia,
  ``General fluctuations of the type II pure spinor string on curved backgrounds,''
JHEP {\bf 1904}, 073 (2019).
[arXiv:1902.02289 [hep-th]].
}

\lref\ChandiaSTA{
  O.~Chandia and B.~C.~Vallilo,
  ``Non-minimal fields of the pure spinor string in general curved backgrounds,''
JHEP {\bf 1502}, 092 (2015).
[arXiv:1412.1030 [hep-th]].
}

\lref\ChandiaIX{
  O.~Chandia,
 ``A Note on the classical BRST symmetry of the pure spinor string in a curved background,''
JHEP {\bf 0607}, 019 (2006).
[hep-th/0604115].
}

\lref\BerkovitsZZ{
  N.~Berkovits and L.~Mazzucato,
  ``Taming the b antighost with Ramond-Ramond flux,''
JHEP {\bf 1011}, 019 (2010).
[arXiv:1004.5140 [hep-th]].
}

\lref\JusinskasVMD{
  R.~L.~Jusinskas,
  ``Towards the underlying gauge theory of the pure spinor superstring,''
JHEP {\bf 1910}, 063 (2019).
[arXiv:1903.10753 [hep-th]].
}

\lref\BerkovitsGH{
  N.~Berkovits,
  ``Pure spinors, twistors, and emergent supersymmetry,''
JHEP {\bf 1212}, 006 (2012).
[arXiv:1105.1147 [hep-th]].
}

\lref\BerkovitsAIA{
  N.~Berkovits,
  ``Twistor Origin of the Superstring,''
JHEP {\bf 1503}, 122 (2015).
[arXiv:1409.2510 [hep-th]].
}

\lref\BerkovitsYRA{
  N.~Berkovits,
  ``Origin of the Pure Spinor and Green-Schwarz Formalisms,''
JHEP {\bf 1507}, 091 (2015).
[arXiv:1503.03080 [hep-th]].
}

\lref\FleuryBSD{
  T.~Fleury,
  ``On the Pure Spinor Heterotic Superstring $b$ Ghost,''
JHEP {\bf 1603}, 200 (2016).
[arXiv:1512.00807 [hep-th]].
}

\newsec{Introduction}

The pure spinor formalism was invented almost twenty years ago to solve the problems arising in the covariant quantization of the Green-Schwarz superstring \BerkovitsFE. Since then, the formalism has passed several tests. The formalism has the correct physical spectrum \BerkovitsNN, \BerkovitsMX, reproduces scattering amplitudes at tree-level \BerkovitsPH\ and at higher loops level \MafraWQ. It is also useful to construct massive states in a covariant language  \BerkovitsQX. The pure spinor formalism was used to couple a string to a curved background \BerkovitsUE\ including background with Ramond-Ramond background fields like the ones present in the $AdS_5\times S^5$ background geometry for the type IIB superstring \BerkovitsYR. 

One of the more mysterious aspects of the pure spinor string is the absence of the world-sheet reparametrization ghosts as fundamental fields. They come from gauge-fixing the reparametrization invariance of the traditional string theories. In the conventional conformal gauge, they are the left-moving $(b_L, c_L)$ and right-moving $(b_R, c_R)$ ghosts. It is not known if the pure spinor string comes from gauge fixing a set of local world-sheet symmetries that includes the reparametrization symmetry (although this problem was recently discussed in \JusinskasVMD), then the reparametrization ghosts do not appear in a BRST 	quantization scheme in a natural way. However, the $b$ ghost is a necessary ingredient to construct higher-loops scattering amplitudes \BerkovitsZZ. Although there is no $c$ ghosts in the pure spinor formalism\foot{See \BerkovitsGH, \BerkovitsAIA, \BerkovitsYRA\ for discussions about this issue.}, the $b$ ghosts are constructed as a composite field and after noticing that the stress-energy tensor $T$ is BRST closed and trivial in flat spacetime background \BerkovitsBT, in $AdS_5\times S^5$ background \BerkovitsZZ\ and in a generic supergravity background for the heterotic string \BerkovitsAMA. In this way, the $b$ ghost is defined to satisfy $Qb=T$. A natural generalization is the case of the type II superstring in a generic curved background. This is the goal of this work. 

To find a conformal weight two an ghost number zero satisfying $Qb=T$ it is necessary the introduction of new variables. They are the so called non-minimal pure spinor variables \BerkovitsBT. These new constrained variables do not change the physical content of the pure spinor string and help to construct the $b$ ghost field. In this way, the $b$ ghost in flat spacetime background was constructed in \BerkovitsBT. The construction was simplified in \BerkovitsPLA. An obvious generalization is to turn on background supergravity fields. It is known that such superfields are constrained by the BRST symmetry \BerkovitsUE. For the heterotic string in curved background, the non-minimal variables have a non trivial coupling to the background and the BRST transformations depend on background torsion superfields \ChandiaIMA. As in flat spacetime \BerkovitsPLA, the construction of the $b$ ghost is simplified because the existence of a RNS-like vector \BerkovitsAMA. The goal of this paper is to first find the coupling of the non-minimal variables to type II background supergeometry  and then construct the $b$ ghosts. 

The plan of this paper is as follows. In section 2 we review the non-minimal type II pure spinor string in flat background. In section 3 we review the minimal pure spinor formalism for the type II superstring in a curved background. In section 4 we study the non-minimal pure spinor variables in a curved background. In section 5 we construct the $b$ ghosts for the type II superstring. 

\newsec{Non-minimal type II pure spinor  superstring in a flat background}

The action of the non-minimal type II superstring in a flat background has the form
\eqn\sz{\eqalign{S&=\int d^2z \ha\p X_m \pb X^m + p_{L\a} \pb\t_L^\a+p_{R\ab}\p\t_R^\ab+\o_{L\a}\pb\l_L^\a+\o_{R\ab}\p\l_R^\ab \cr&+\oh_L^\a \pb\lh_{L\a}+s_L^\a\pb r_{L\a}+ \oh_R^\ab \p \lh_{R\ab}+s_R^\ab\p r_{R\ab} , \cr}}
where $(X^m,\t_L^\a,\t_R^\ab)$ are the coordinates of N=2 ten-dimensional superspace ($m=0, \dots, 9; \a,\ab=1, \dots, 16$). The chiralities $\a$ and $\ab$ are the opposite for type IIA superstring and are the same for type IIB superstring. The momentum conjugate variables of $(\t_L^\a,\t_R^\ab)$ are $(p_{L\a}, p_{R\ab})$.  The pure spinor variables $(\l_L,\l_R)$ are constrained to satisfy the pure spinor conditions $\l_L\g^m\l_L=\l_R\g^m\l_R=0$ where $\g^m$ are the symmetric $16\times 16$ gamma matrices in ten dimensions. The conjugate variables of $(\l_L^\a, \l_R^\ab)$ are $(\o_{L\a}, \o_{R\ab})$ which are defined up to the pure spinor gauge invariances $\d\o_{L\a}=(\l_L\g^m)_\a \L^1_{Lm}$ and $\d\o_{R\ab}=(\l_R\g^m)_\ab \L^1_{Rm}$. The non-minimal pure spinor variables $r_{L/R}, \lh_{L/R}$ are constrained as
\eqn\nmc{\lh_L\g^m\lh_L=\lh_R\g^m\lh_R=\lh_L\g^m r_L=\lh_R\g^m r_R=0 ,}
and $s_{L/R}, \oh_{L/R}$ are defined up to
\eqn\gauge{\eqalign{&\d s_L^\a=(\g^m\lh_L)^\a \L^2_{Lm},\quad \d\oh_L^\a=(\g^m\lh_L)^\a\L^3_{mL}-(\g^m r_L)^\a \L^2_{Lm} \cr&\d s_R^\ab=(\g^m\lh_R)^\ab \L^2_{Rm},\quad \d\oh_R^\ab=(\g^m\lh_R)^\ab\L^3_{mR}-(\g^m r_R)^\ab \L^2_{Rm} .\cr }} 
The pure spinor BRST charge of the system is defined by $Q=Q_L+Q_R$, where
\eqn\Qf{\eqalign{&Q_L=\oint( \l_L^\a d_{L\a} + \oh_L^\a r_{L\a} ) ,\cr
&Q_R=\oint(\l_R^\ab d_{R\ab}+\oh_R^\ab r_{R\ab} ) ,\cr}}
and
\eqn\df{\eqalign{&d_{L\a}=p_{L\a}-\ha(\g^m\t_L)_\a(\p X_m+{1\over4}(\t_L\g_m\p\t_L)),\cr&d_{R\ab}=p_{R\ab}-\ha(\g^m\t_R)_\ab(\pb X_m+{1\over4}(\t_R\g_m\pb\t_R)) .\cr }}
Note that these variables satisfy the OPE algebra
\eqn\dope{\eqalign{&d_{L\a}(y) d_{R\ab}(\zz)\to 0 ,\cr &d_{L\a}(y) d_{L\b}(z) \to -{1\over(y-z)}\g^m_{\a\b}\Pi_m ,\cr & d_{R\ab}(\yy) d_{R\bb}(\zz) \to -{1\over(\yy-\zz)}\g^m_{\ab\bb}\Pib_m ,\cr }}
where
\eqn\pis{\Pi_m=\p X_m+\ha(\t_L\g_m\p\t_L),\quad \Pib_m=\pb X_m+\ha(\t_R\g_m\pb\t_R) .}
Using the free-field OPE algebra of the non-minimal pure spinor variables,  the algebra \dope\ and the pure spinor conditions \nmc\ one can show that $Q_L^2=Q_R^2=\{Q_L,Q_R\}=0$. This imply that $Q=Q_L+Q_R$ is nilpotent. Note also that the second integrands in \Qf, the non-minimal contribution to the BRST charge, have trivial cohomology \BerkovitsBT. 

Consider the transformations generated by $Q_L$. Acting on the minimal variables $(\l_L^\a, \t_L^\a, \Pi^m, d_{L\a}, \o_{L\a}), Q_L$ gives
\eqn\QLf{Q_L\l_L^\a=0,\quad Q_L\t_L^\a=\l_L^\a,\quad Q_L\Pi^m=\l_L\g^m\p\t_L,\quad Q_L d_{L\a}=-(\l_L\g_m)_\a \Pi^m,\quad Q\o_{L\a}=d_{L\a} ,}
and acting on the non-minimal variables $(\lh_{L\a}, \oh_L^\a, r_{L\a}, s_L^\a), Q_L$ gives
\eqn\QLnm{Q_L\lh_{L\a}=-r_{L\a},\quad Q_L\oh_L^\a=0,\quad Q_L r_{L\a}=0,\quad Q_L s_L^\a=\oh_L^\a .}
 Note that $Q_R$ annihilates both set of variables. 

The stress-energy tensor 
\eqn\TLf{T_L=-\ha\Pi_m\Pi^m-d_{L\a}\p\t_L^\a-\o_{L\a}\p\l_L^\a-\oh_L^\a\p\lh_{L\a}-s_L^\a\p r_{L\a} ,} 
is BRST invariant. As in \BerkovitsBT, one can ask if there exists a conformal dimension two $b_L$ satisfying
\eqn\Qbf{Q_L b_L=T_L .} 
The answer is yes. A way to get $b_L$ is to first define a RNS-like vector 
\eqn\GLf{\Gb_L^m={1\over{2(\l_L\lh_L)}}(d_L\g^m\lh_L)-{1\over{8(\l_L\lh_L)^2}}(r_L\g^{mnp}\lh_L)N_{np} ,}
where  $N_{np}=\ha(\l_L\g_{np}\o_L)$, that transforms under $Q_L$ as 
\eqn\QLGLf{Q\Gb_L^m=-{1\over{2(\l_L\lh_L)}}\Pi^n(\lh_L\g^m\g_n\l_L)+{1\over{4(\l_L\lh_L)^2}}(\l_L\g^{np}r_L)(\lh_L\g_p\g^m\l_L)\Gb_{Ln} ,}
and $Q_R\Gb_L^m=0$. A $b_L$ satisfying \Qbf\ is given by
\eqn\bLf{b_L=-s_L^\a\p\lh_{L\a}-\o_{L\a}\p\t_L^\a+\Pi_m\Gb_L^m+{1\over{4(\l_L\lh_L)}}(\l_L\g_{mn}r_L)\Gb_L^m\Gb_L^n+{1\over{2(\l_L\lh_L)}}(\o_L\g^m\lh_L)(\l_L\g_m\p\t_L) ,}
and it also satisfies $Q_Rb_L=0$. The proofs of  \QLGLf\ and \Qbf\  are easy to do after noting that the transformations generated by the type II BRST charge $Q_L$ have the same form that the transformations generated by the heterotic BRST charge in  \BerkovitsAMA.  

A similar analysis can be done from $Q_R$. In this case the relevant world-sheet variables are the minimal pure spinor variables $(\l_R^\ab, \t_R^\ab, \Pib^m, d_{R\ab}, \o_{R\ab})$ as well as the non-minimal pure spinor variables $(\lh_{R\ab}, \oh_R^\ab, r_{R\ab}, s_R^\ab)$. The BRST transformations are given by $Q_R$ and $Q_L$ annihilates these variables. It turns out that there is a $b_R$ field satisfying $Q_R b_R=T_R$ and $Q_L b_R=0$ where $T_R$ is the right-moving stress-energy tensor.

Our purpose is to generalize the analysis of this section for a type II string in a curved background. 

\newsec{The minimal type II pure spinor string in a curved background}

The action for the type II superstring in the minimal pure spinor formalism is
\eqn\Sc{\eqalign{S_0&=\int d^2z ( ~\ha \Pi_a \Pib^a + \ha \Pi^A \Pib^B B_{BA}+d_{L\a} \Pib^\a + d_{R\ab} \Pi^\ab+\o_{L\a}\Nb\l_L^\a+\o_{R\ab}\N\l_R^\ab \cr&+d_{L\a}d_{R\ab}P^{\a\ab}+\l_L^\a\o_{L\b}d_{R\ab} C_\a{}^{\b\ab}+d_{L\a}\l_R^\ab\o_{R\bb}C_\ab{}^{\bb\a}+\l_L^\a \o_{L\b}\l_R^\ab\o_{R\bb}S_{\a\ab}{}^{\b\bb} ) ,\cr}}
where the curved superspace coordinates $Z^M=(X^m, \t_L^\mu, \t_R^{\overline\mu})$ are related to the locally flat superspace coordinates $Z^A=(x^a, \t_L^\a, \t_R^\ab)$ as $Z^A=Z^M E_M{}^A(Z)$, where $E_M{}^A$ is the vielbein superfield. The $\Pi$ world-sheet fields in the action are defined as $\Pi^A=\p Z^M E_M{}^A$. The covariant derivatives in the action are given by
\eqn\covD{\Nb\l_L^\a=\pb\l_L^\a+\l_L^\b\pb Z^M \O_{M\b}{}^\a, \quad \N\l_R^\ab=\p\l_R^\ab+\l_R^\bb\O_{M\bb}{}^\ab ,}
where the $\O$  superfields are the connections for the background symmetries. They are defined as 
\eqn\Lor{\O_{M\a}{}^\b=\d_\a^\b\O_M+{1\over4}(\g^{ab})_\a{}^\b\O_{Mab} ,\quad \O_{M\ab}{}^\bb=\d_\ab^\bb\Oh_M+{1\over4}(\g^{ab})_\ab{}^\bb\Oh_{Mab} .}
Here $\O_M$ and $\Oh_M$ are the connections for the scaling symmetry of the action \Sc, while $\O_{Mab}$ and $\Oh_{Mab}$ are connections for the Lorentz rotation symmetry of the action \Sc. 

The BRST symmetry for the minimal pure spinor string is generated by 
\eqn\Qmin{Q_0=Q_{0L}+Q_{0R}=\oint \l_L^\a d_{L\a} + \oint \l_R^\ab d_{R\ab} ,}
which is nilpotent and conserved when the background fields satisfy the type II supergravity equations of motion in ten dimensions. The constraints are expressed in terms of the torsion, curvature and the $3$-form $H=dB$ components as can be seen in \BerkovitsUE. The torsion and the curvature $2$-forms are defined as
\eqn\TRis{T^A=\N E^A=dE^A+E^B\O_B{}^A,\quad R_A{}^B=d\O_A{}^B+\O_A{}^C\O_C{}^A ,} 
and the constraints from BRST invariance are solved by 
\eqn\solTR{\eqalign{&T_{\a\b a}=H_{\a\b a}=-(\g_a)_{\a\b}, \quad T_{\ab\bb a}=-H_{\ab\bb a}=-(\g_a)_{\ab\bb},\quad T_{a\a}{}^\b=T_{a\ab}{}^\bb=0,\cr&R_{\a\b\gb}{}^\db=R_{\ab\bb\g}{}^\d=H_{A\a\bb}=H_{ab\a}=H_{ab\ab}=0    ,\cr}}
and $T_{AB}{}^C=H_{ABC}=0$ whenever $(A, B, C)\in (\a,\ab)$. 

The torsion and curvature $2$-forms satisfy the Bianchi identities
\eqn\bianchi{\N T^A=T^B R_B{}^A,\quad \N R_A{}^B=0. }
We have to be careful here because, as it was mentioned above, when the index $A$ takes the value $a$ there are two possible connections to be used, $\O^{ab}$ or $\Oh^{ab}$ of \Lor. The analysis of the Bianchi identities taking care of this subtlety can be found in \BedoyaIC\ (see also \ChandiaPAJ) we just quote the result
\eqn\torsions{\eqalign{&T_{\a a}{}^b=2(\g_a{}^b)_a{}^\b\O_\b,\quad T_{\ab a}{}^b=\O_\ab=\O_a=0,\cr&\Th_{\ab a}{}^b=2(\g_a{}^b)_\ab{}^\bb\Oh_\bb,\quad \Th_{\a a}{}^b=\Oh_\a=\Oh_a=0,\cr&T_{abc}=-H_{abc},\quad \Th_{abc}=H_{abc} .\cr}}
The components of the connection $1$-form $\Oh^{ab}$ are given in terms of the connection $1$-form $\O^{ab}$ and torsion components as
\eqn\Ohis{\Oh_c{}^{ab}=\O_c{}^{ab}-T_c{}^{ab},\quad \Oh_\a{}^{ab}=\O_\a{}^{ab}-T_\a{}^{ab},\quad \Oh_\ab{}^{ab}=\O_\ab{}^{ab}+\Th_\ab{}^{ab} ,}
where the local vector indices are raised or lowered with $\eta^{ab}$ or $\eta_{ab}$ respectively. 

The other background superfields in the action \Sc\ satisfy  
\eqn\other{\eqalign{&T_{a\a}{}^\ab=(\g_a)_{\a\b}P^{\b\ab},\quad R_{a\a\ab}{}^\bb=-(\g_a)_{\a\b} C_\ab{}^{\bb\b},\cr&C_\a{}^{\b\ab}=-\N_\a P^{\b\ab},\quad
S_{\a\ab}{}^{\b\bb}=\N_\a C_\ab{}^{\bb\b}+R_{\a\gb\ab}{}^\bb P^{\b\gb},\cr&T_{a\ab}{}^\a=-(\g_a)_{\ab\bb}P^{\a\bb},\quad R_{a\ab\a}{}^\b=-(\g_a)_{\ab\bb}
C_\a{}^{\b\bb},\cr&C_\ab{}^{\bb\a}=\N_\ab P^{\a\bb},\quad S_{\a\ab}{}^{\b\bb}=\N_\ab C_\a{}^{\b\bb}-R_{\ab\g\a}{}^\b P^{\g\bb}  .\cr }}

The BRST transformations generated by $Q_{L0}$ was determined in \ChandiaSTA. The variables $(\l_L^\a, \o_{L\a}, d_{L\a})$ transform as
\eqn\QLz{\eqalign{&Q_{L0}\l_L^\a=-\l_L^\b \S_{L\b}{}^\a,\quad Q_{L0}\o_{L\a}=d_{L\a}+\S_{L\a}{}^\b\o_{L\b} ,\cr
&Q_{L0} d_{L\a}=-(\l_L\g_a)\Pi^a + \l_L^\b R_{\a\b\g}{}^\d \l_L^\g\o_{L\d}+\S_{L\a}{}^\b d_{L\b} ,\cr}}
where $\S_{L\a}{}^\b=\l_L^\g\O_{\g\a}{}^\b$ is a field-dependent Lorentz transformation parameter. Note that from \Lor\ 
\eqn\Sis{\S_{L\a}{}^\b=\d_\a^\b\S_L+{1\over4}(\g^{ab})_\a{}^\b \S_{Lab} ,}
where $\S_L=\l^\g\O_\g$ and $\S_{Lab}=\l^\g\O_{\g ab}$. The appearance of a Lorentz rotation term in the BRST transformations is usual in curved backgrounds and it was first noted in \ChandiaIX. In fact, for $\Pi^a$ and $\Pi^\a$ one obtains from $Q_{L0}Z^M=\l_L^\a E_\a{}^M$ that they transform as
\eqn\QLzPi{Q_{L0}\Pi^\a=\N\l_L^\a+\Pi^\b\S_{L\b}{}^\a,\quad Q_{L0}\Pi^a=\l_L^\a\Pi^\b \g^a_{\a\b}+\l_L^\a\Pi^b T_{\a b}{}^a-\Pi^b\S_{Lb}{}^a ,}
where the covariant derivative is defined in \covD.

As in flat space, the left-moving stress energy tensor 
\eqn\TLc{T_L=-\ha\Pi_a\Pi^a-d_{L\a}\Pi^\a-\o_{L\a}\N\l_L^\a ,}
is annihilated by $Q_{L0}$. The proof of this requires the BRST transformation of $\Pi^A \O_{A\a}{}^\b$ inside the covariant derivative in \TLc. It turns out that
\eqn\QLO{Q_{L0}(\Pi^A \O_{A\a}{}^\b)=-\l^\g\Pi^A R_{A\g\a}{}^\b+\N\S_{L\a}{}^\b .}
Again, here appears a Lorentz rotation term. Using the above transformations it is direct to verify that $Q_{L0}T_L$ vanishes. As in flat spacetime, one defines the ghost $b_L$ such that $Q_{L0}b_L=T_L$. But, the ghost $b_L$ will need the presence of the non-minimal pure spinor variables that are introduced in the next section.

The transformations of the left-moving minimal pure spinor variables under $Q_{R0}$ are
\eqn\QRonL{\eqalign{&Q_{R0}\l_L^\a=-\l_L^\b \S_{R\b}{}^\a,\quad Q_{R0}\o_{L\a}=\S_{R\a}{}^\b\o_{L\b} ,\cr
&Q_{R0} d_{L\a}= \l_R^\bb R_{\a\bb\g}{}^\d \l_L^\g\o_{L\d}+\S_{R\a}{}^\b d_{L\b} ,\cr}}
where $\S_{R\a}{}^\b=\l_R^\gb\O_{\gb\a}{}^\b$ is a field-dependent Lorentz rotation parameter which is expressed like $\S_{R\a}{}^\b={1\over4}(\g^{ab})_\a{}^\b\S_{Rab}$ where $\S_{Rab}=\l_R^\gb\O_{\gb ab}$. Note that, unlike \Sis, there is no term with $\d_\a^\b$ because $\O_\gb=0$. Except for the term with curvature, the transformations \QRonL\ are Lorentz rotations. Similarly, the transformations of $\Pi^\a$ and $\Pi^a$ under $Q_{R0}$ are determined from $Q_{R0}Z^M=\l_R^\ab E_\ab{}^M$ and are given by
\eqn\QRPIL{Q_{R0}\Pi^\a=\Pi^a(\l_R\g_a P)^\a-\Pi^\b \S_{R\b}{}^\a,\quad Q_{R0}\Pi^a=\l_R^\ab\Pi^\bb\g^a_{\ab\bb}-\Pi^b\S_{Rb}{}^a ,}
where a Lorentz rotation part is manifest.  

\newsec{The introduction of the non-minimal pure spinor variables in a curved background}

The non-minimal pure spinor variables are given by the left moving sector $(\lh_{L\a}, \oh_L^\a, r_{L\a}, s_L^\a)$ and the right-moving sector $(\lh_{R\ab}, \oh_R^\ab, r_{R\ab}, s_R^\ab)$. The BRST charge of the left-moving variables is
\eqn\QLone{Q_{L1}=\oint \oh_L^\a r_{L\a} ,} 
just like flat space-time \Qf. Similarly, for the right moving sector
\eqn\QRone{Q_{R1}=\oint \oh_R^\ab r_{R\ab} .} 
As in \ChandiaIMA, we asume that both $Q_{L1}$ and $Q_{R1}$ act on the non-minimal pure spinor variables just like in flat spacetime. The question what is $Q_0$ on these variables. Since the non-minimal variables transform under Lorentz transformations, they are expected to transform under BRST because it always generates Lorentz rotations. Consider $Q_L=Q_{L0}+Q_{L1}$. We propose that the left-moving non-minimal transformations are
\eqn\QLnm{\eqalign{&Q_L\lh_{L\a}=-r_{L\a}+\St_{L\a}{}^\b\lh_{L\b},\quad Q_L\oh_L^\a=-\oh_L^\b\St_{L\b}{}^\a ,\cr
&Q_L s_L^\a=\oh_L^\a+s_L^\b\St_{L\b}{}^\a,\quad Q_Lr_{L\a}=\St_{L\a}{}^\b r_{L\b},\cr }}
where $\St_{L\a}{}^\b=\S_{L\a}{}^\b+X_{L\a}{}^\b$ with $X_{L\a}{}^\b=\l_L^\g X_{L\g\a}{}^\b$ not known yet. Because $Q_L$ is nilpotent $\St_L$ must obey certain equation which is obtained after acting with $Q_L$ in the transformations \QLnm. The equation for $\St_L$ turns out to be 
\eqn\eqXL{\l_L^\g\l_L^\d \left( R_{\g\d\a}{}^\b + \N_{(\g} X_{L\d)\a}{}^\b -X_{L(\g\a}{}^\r X_{L\d)\r}{}^\b \right) =0 ,}
which has solution
\eqn\XLis{X_{L\g\a}{}^\b=x_L\d_\a^\b\O_\g-{1\over4}(\g^{ab})_\a{}^\b T_{\g ab} ,}
where $x_L$ is a number not fixed yet. To proof this one expand the lhs of \eqXL\ in $_\a{}^\b$ using the gamma matrices in ten dimensions. The term with $(\g^{abcd})_\a{}^\b$ is absent. The term with $\d_\a^\b$ is proportional to
\eqn\pone{\l_L^\g\l_L^\d ~\N_{(\g}\O_{\d)}={1\over4} \l_L^\g\l_L^\d \{\N_\g,\N_\d\}\Phi={1\over4}(\l_L\g^a\l_L)\N_a\Phi,}
which vanishes because of the pure spinor condition. Here $\Phi$ is the dilaton superfield \BerkovitsUE. The term with $(\g^{ab})_\a{}^\b$ is proportional to
\eqn\ptwo{\l_L^\g\l_L^\d \left( R_{\g\d ab}-\N_{(\g} T_{\d)ab}-T_{(\g a}{}^c T_{\d)cb} \right) ,}
using the Bianchi identity involving $R_{\g\d ab}$ one obtains that \ptwo\ is proportional to
\eqn\pthree{(\l_L\g^c\l_L) T_{c ab} ,}
which vanishes because of the pure spinor condition. Therefore, \XLis\ is solution of \eqXL.

Similarly, for the right-moving non-minimal variables transform under $Q_R=Q_{R0}+Q_{R1}$ as
\eqn\QRnm{\eqalign{&Q_R\lh_{R\ab}=-r_{R\a}+\St_{R\ab}{}^\bb\lh_{R\bb},\quad Q_R\oh_R^\ab=-\oh_R^\bb\St_{R\bb}{}^\ab ,\cr
&Q_R s_R^\ab=\oh_R^\ab+s_R^\bb\St_{R\bb}{}^\ab,\quad Q_Rr_{R\ab}=\St_{R\ab}{}^\bb r_{R\bb},\cr }}
where $\St_{R\ab}{}^\bb=\S_{R\ab}{}^\bb+X_{R\ab}{}^\bb$ with $X_{R\ab}{}^\bb=\l_R^\gb X_{R\gb\ab}{}^\bb$ not known yet. Because $Q_R$ is nilpotent $\St_R$ must obey certain equation which is obtained after acting with $Q_R$ in the transformations \QRnm. The equation for $\St_R$ turns out to be 
\eqn\eqXR{\l_R^\gb\l_R^\db \left( R_{\gb\db\ab}{}^\bb + \N_{(\gb} X_{R\db)\ab}{}^\bb -X_{R(\gb\ab}{}^\rb X_{R\db)\rb}{}^\bb \right) =0 ,}
which has solution
\eqn\XRis{X_{R\gb\ab}{}^\bb=x_R\d_\ab^\bb\Oh_\gb-{1\over4}(\g^{ab})_\ab{}^\bb \Th_{\gb ab} ,}
where $x_R$ is a number not fixed yet. 

\newsec{The RNS-like vector and the $b$ ghost}

We now mix minimal and non-minimal pure spinor variables to construct the $b$ ghost fields for type II backgrounds. As in \BerkovitsAMA\ for the heterotic case, we first define variables with BRST transformations similar to those of flat space-time. Consider the left-moving variables. The right-moving sector will have a similar behavior. The non-minimal ghosts transform under $Q_L$ with expressions that are the addition of the corresponding transformation in flat space-time and a Lorentz rotation parametrized by $\St_L$ \QLnm. We now define a variable in  the minimal sector that has the same property. The new variable involving  $d_L$ is
\eqn\newDL{D_{L\a}=d_{L\a}-X_{L\a}{}^\b\o_{L\b} ,}
where $X_L$ is used in \QLnm\ to define the left-moving BRST transformations of the non-minimal pure spinor ghosts. Using the above results, $D_L$ transforms as
\eqn\QLDL{Q_L D_{L\a}=-(\l_L\g_a)_\a \Pi^a + \St_{L\a}{}^\b D_{L\b} ,}
which has the desired property, a rotation and a flat-like part.

Since $D_L$ involves the pure spinor $\o_L$ variable, the rhs of \newDL\ has to be invariant under $\d\o_{L\a}=(\l_L\g^a)_\a \L_a$. Using the solution of $X_L$ given in \XLis, this is true provided $x_L=3$. Following \BerkovitsAMA, the RNS-like vector $\Gb_{La}$ is the the flat expression \GLf\ by replacing $d_L$ by $D_L$. Then,
\eqn\GBL{\Gb_{La}={1\over{2(\l_L\lh_L)}} (D_L\g_a\lh_L)-{1\over{8(\l_L\lh_L)^2}}(r_L\g_{abc}\lh_L) N^{bc} ,}
and it transforms like
\eqn\QGbhc{Q_L\Gb_{La}=-{1\over{2(\l_L\lh_L)}}\Pi_b (\lh_L\g_a\g^b\l_L)+{1\over{4(\l_L\lh_L)^2}}(\l_L\g^{bc}r_L)(\lh_L\g_c\g_a\l_L)\Gb_{Lb} + \St_{La}{}^b \Gb_{Lb} .}
This is equivalent to the corresponding BRST transformation in flat space-time \QLGLf\ expect for the Lorentz rotation in the last term. To prove \QGbhc, we need to express the BRST transformations of the fields in \GBL\ as a rotation parametrized by $\St$ plus a term that has the form of the corresponding BRST transformation in flat space-time. The fields $(D_L, \lh_L, r_L)$ already transform in this way. It remains to express the BRST transformations of $(\l_L, \o_L)$ in this way. Consider $\l_L$, 
$$Q_L\l_L^\a=-\l_L^\b\S_{L\b}{}^\a=-\l^\b(\St_{L\b}{}^\a-X_{L\b}{}^\a) ,$$
and using \XLis\ with $x_L=3$ we obtain
\eqn\Qlnew{Q_L\l_L^\a=8(\l_L\O)\l_L^\a-\l_L^\b\St_{L\b}{}^\a ,}
which has the form of the flat space-time form, zero, plus a rotation parametrized by $\St$ and an extra term with the factor $(\l_L\O)$. Consider now $\o_L$, 
$$Q_L\o_{L\a}=d_{L\a}+\S_{L\a}{}^\b\o_{L\b}=D_{L\a}+(\S_{L\a}{}^\b+X_{L\a}{}^\b)\o_{L\b},$$
that is,
\eqn\Qonew{Q_L\o_{L\a}=D_{L\a}+\St_{L\a}{}^\b\o_\b ,}
which has the expected form.

Note that the world-sheet fields $D_L, \Gb_L$ and $\o_L$ transforms under $Q_L$ as the corresponding transformation in flat space-time plus a Lorentz rotation parametrized by $\St_L$. This i simpler than  the result of \BerkovitsAMA\ for the heterotic string. I believe the result of this paper is a small improvement of the result of \BerkovitsAMA\ such that computations that involve BRST transformations of Lorentz invariant quantities, like construction of the $b_L$ ghost, are easier to perform.

In order to prove \QGbhc\ we need to verify that all the factors of $(\l_L\O)$ cancel out. These factors came from the BRST transformation of $\l_L$ and the part of Lorentz rotation $\St_L$ with the factor of $\d_\a^\b$. That is, combining \Sis\ and \XLis\ we obtain
\eqn\Sthis{\St_{L\a}{}^\b=\d_\a^\b\St_L+{1\over4}(\g^{ab})_\a{}^\b\St_{Lab}=4\d_\a^\b(\l_L\O)+{1\over4}(\g^{ab})_\a{}^\b\l_L^\g(\O_{\g ab}-T_{\g ab}).}
It turns out that both terms in \GBL\ do not give term with $(\l_L\O)$ factors. Consider the first term in \GBL\ and focus on $(\l_L\O)$ factors,
\eqn\QGone{Q_L\left({1\over{2(\l_L\lh_L)}}(D_L\g_a\lh_L)\right)={1\over{(\l_L\lh_L)}} (-4(\l_L\O)+\St_L)(D_L\g_a\lh_L)+\cdots=0+\cdots ,}
where we used $\St_L=4(\l_L\O)$ \Sthis. The first term here comes from transforming $(\l_L\lh_L)^{-1}$ and the second term comes from the part with $\St_L$ in the transformations of $D_{L\a}$ and $\lh_{L\a}$ in the numerator of the first term in \GBL. The factor of $\lh$ in the denominator do not give factors of $\St_L$ because it appears in a Lorentz invariant combination. Similarly, the for second term in \GBL\ we obtain
\eqn\QGtwo{\eqalign{Q_L\left( {1\over{8(\l_L\lh_L)^2}}(r_L\g_{abc}\lh_L)N^{bc} \right)&={1\over{(\l_L\lh_L)^2}} \left( -2(\l_L\O)+{1\over4}\St_L+(\l_L\O) \right) (r_L\g_{abc}\lh_L) N^{bc}+\cdots\cr&=0+\cdots ,\cr}}
where the first term comes from the transformation of $(\l_L\lh_L)^{-2}$, the second term comes from the transformations of $r_L$ and $\lh_L$, and the last term comes from $\l_L$ in $N^{bc}$. Therefore, we have proved \QGbhc. The construction of $b_L$ ghost will be simpler than the corresponding construction in \BerkovitsAMA\ because this ghost is a Lorentz invariant quantity. 

The $b_L$ ghost satisfies $Q_Lb_L=T_L$, then we need to know the stress-energy tensor of the theory. As in  \ChandiaIMA, the action is
\eqn\Shc{S=S_0+Q_L\left( \int d^2z ~s_L^\a\Nb\lh_{L\a} \right) + \cdots,}
where $S_0$ is given in \Sc\ and $\cdots$ is the contribution from the right-moving non-minimal pure spinor variables. The left-moving stress-energy tensor becomes
\eqn\Thet{T_L=-\ha\Pi_a\Pi^a-d_{L\a}\Pi^\a-\o_{L\a} \N\l_L^\a-Q_L\left(s_L^\a\N\lh_{L\a}\right) .
}
The $b_L$ ghost is the covariantization of \bLf, that is
\eqn\bhc{b_L=-s_L^\a\N\lh_{L\a}-\o_{L\a}\Pi^\a+\Pi^a\Gb_{La}+{1\over{4(\l_L\lh_L)}}(\l_L\g^{ab}r_L)\Gb_{La}\Gb_{Lb}+{1\over{2(\l_L\lh_L)}}(\o_L\g^a\lh_L)(\l_L\g_a\Pi).
}
To prove $Q_Lb_L=T_L$ we need to re-express the $Q_L$ BRST transformations of $(\Pi^a, \Pi^\a)$ \QLzPi\ such that they contain a Lorentz rotation with parameter $\St_L$. Using \QLzPi\ we obtain
\eqn\QPihnew{Q_L\Pi^a=(\l_L\g^a\Pi)-\Pi^b\St_{Lb}{}^a,\quad Q_L\Pi^\a=\N\l_L^\a+\Pi^\b\St_{L\b}{}^\a+3(\l_L\O)\Pi^\a-\ha(\l_L\g_{ab}\O)(\Pi\g^{ab})^\a,}
where we note, up to terms with $\O_\a$, the structure of a flat-like part plus a rotation of the BRST transformations. In the computation of $Q_Lb_L$ we note that the first term gives the last term in \Thet. The remaining terms in $b_L$ should give the remaining terms of \Thet. Because of the form of the BRST transformations, this is achieved except for terms with factors $(\l_L\O)$. These comes from $(\l_L\lh_L)^{-1}$ and the scalar part in the $Q_L$ BRST transformations of $\o_{L\a}$ together with the $Q_L$ BRST transformations of $\Pi^\a$. The terms with factor of $(\l_L\lh_L)^{-1}$ are with a factor of $\l_L^\a$, this combination does not contain $(\l_L\O)$. In fact, ${\l_L^\a/{(\l_L\lh_L)}}$ produces
$$-{1\over{(\l_L\lh_L)^2}} 8(\l_L\O)(\l_L^\b\lh_{L\b})\l_L^\a+{1\over{(\l_L\lh_L)}}8(\l_L\O)\l_L^\a=0.$$
When transforming the second term of the $b_L$ ghost one produces $-d_{L\a}\Pi^\a$, which is part of the stress-energy tensor $T_L$, but one also obtains terms involving the $X_L$ of $\St_L$. In fact we get the combination
$$(X_{L\a}{}^\b \o_{L\b}) \Pi^\b+\o_{L\a}(\Pi^\b X_{L\b}{}^\a),$$
which vanishes. The first term comes from the term with $X_L$ in \newDL\ and the second term is made from the last two terms in $Q\Pi^\a$ of \QPihnew. It remains to check that the terms with $\O_\a$ go away when the $Q_L$ BRST charge acts on the last term of the $b_L$ ghost \bhc. The term with $D_L$ in $Q_L\o_L$ here will help to prove the equation $Q_Lb_L=T_L$, just like in flat space-time (see the appendix of \BerkovitsAMA). The only non-trivial contribution, and that has to vanish, comes from the terms with $\O_\a$ in $Q_L\Pi^\a$. They give a term involving the factor
\eqn\qbth{(\l_L\g_a)_\a\Pi^\b\left(3\d_\b^\a(\l_L\O)-\ha(\l_L\g_{bc}\O)(\g^{bc})_\b{}^\a\right)=3(\l_L\g_a\Pi)(\l_L\O)+\ha(\l_L\g_a\g_{bc}\Pi)(\l_L\g^{bc}\O).}
After commuting the $\g$ matrices in the last term, using the identity $(\l\g_b)_\a(\l\g^b)_\b=0$ and $\g_b\g_a\g^b=-8\g_a$, we find that \qbth\ vanishes. This completes the proof of $Q_Lb_L=T_L$. 

Consider the right-moving part. It is possible to define a combination of minimal variables in the right-moving sector analogous to \newDL\ as
\eqn\newDR{D_{R\ab}=d_{R\ab}-X_{R\ab}{}^\bb\o_{R\bb} ,}
where $X_R$ is given in the $Q_R$ transformation of the right-moving non-minimal pure spinor variables \QRnm. Acting with $Q_R$ on $D_R$ one obtains
\eqn\QRDR{Q_R D_{R\ab}=-(\l_R\g_a)_\ab \Pib^a+\St_{R\ab}{}^\bb D_{R\bb} ,}
which is analogous to \QLDL. The world-sheet field $D_R$ is invariant under the pure spinor gauge symmetry $\d\o_{R\a}=(\l_R\g_a)_\ab{\overline\L}^a$ as long as the $x_R$ of \XRis\ is equal to $3$. The RNS-like vector becomes 
\eqn\GBR{\Gb_{Ra}={1\over{2(\l_R\lh_R)}}(D_R\g_a\lh_R)-{1\over{8(\l_R\lh_R)^2}}(r_R\g_{abc}\lh_R)\NN^{bc} ,}
which transforms as
\eqn\QRGB{Q_R\Gb_{Ra}=-{1\over{2(\l_R\lh_R)}}\Pib_b(\lh_R\g_a\g^b\l_R)+{1\over{4(\l_R\lh_R)^2}}(\l_R\g^{bc}r_R)(\lh_R\g_c\g_a\l_R)\Gb_{Rb}+\St_{Ra}{}^\b\Gb_{Rb} ,}
which is the analogous of \QGbhc\  for the left-moving sector. The $b_R$ ghost satisfying $Q_Rb_R=T_R$ where
$$T_R=-\ha\Pib_a\Pib^a-d_{R\ab}\Pib^\ab-\o_{R\ab}\Nb\l_R^\ab-Q_R\left(s_R^\ab\Nb\lh_{R\ab}\right)$$
is given by
\eqn\bR{b_R=-s_R^\ab\Nb\lh_{R\ab}-\o_{R\ab}\Pib^\ab+\Pib^a\Gb_{Ra}+{1\over{4(\l_R\lh_R)}}(\l_R\g^{ab}r_R)\Gb_{Ra}\Gb_{Rb}+{1\over{2(\l_R\lh_R)}}(\o_R\g^a\lh_R)(\l_R\g_a\Pib) .}

It remains to show that the $b$ ghosts satisfy
\eqn\extra{Q_L b_R=0,\quad  Q_R b_L = 0 ,}
in order to have $Qb=T$, where $b=b_L+b_R$ and $T=T_L+T_R$. This is obtained after knowing that non-minimal fields $\lh_L$ and $r_L$ transform under the action of $Q_R$ as a purely rotational part, that is
\eqn\QRlr{Q_R\lh_{L\a}=\S_{R\a}{}^\b\lh_{L\b},\quad Q_R r_{L\a}=\S_{R\a}{}^\b r_{L\b} ,}
and that the non-minimal fields $\lh_R$ and $r_R$ transform under the action of $Q_L$ as a purely rotational part, that is
\eqn\QLlr{Q_L\lh_{R\ab}=\S_{L\ab}{}^\bb\lh_{R\bb},\quad Q_L r_{R\ab}=\S_{L\ab}{}^\b r_{R\bb} .}
Also the transformations $Q_R D_L$ and $Q_L D_R$ are needed. They are
\eqn\QRD{\eqalign{&Q_R D_{L\a}=(\l_L\g^a)_\a(\o_L P\g_a\l_R)-\ha(\l_L\g_a P \g_b\l_R)  (\g^b\g^a\o_L)_\a+\S_{R\a}{}^\b D_{L\b} ,\cr &Q_L D_{R\ab}=-(\l_R\g^a)_\ab(\l_L \g_a P\o_R)+\ha(\l_L\g_aP\g_b\l_R)  (\g^a\g^b\o_R)_\ab+\S_{L\ab}{}^\bb D_{R\bb}  ,\cr}}
where $P$ is the background RR field-strength. Note that these expressions are invariant under the pure spinor gauge symmetries $\d\o_{L\a}=(\l_L\g^a)_\a \L_a$ and $\d\o_{R\a}=(\l_R\g^a)_\ab {\overline\L}_a$ . To proof this one has to use the identities $(\l_L\g^a)_\a(\l_L\g_a)_\b=(\l_R\g^a)_\ab(\l_R\g_a)_\bb=0$. To get \QRD\ the identity$(\g^a P \g_a)_{\a\bb}=8\N_\bb\O_\a$ has been used. It is derived from the Bianchi identity involving $R_{(\bb\a\g)}{}^\g$ and the torsion components in \solTR\ and \other.  

Instead of verify directly the equation \extra, an argument is now given respect to its plausibility. Above It was shown that $Q_L  b_L=T_L$. Applying $Q_R$ and using that it anti-commutes with $Q_L$, 
\eqn\UT{Q_L Q_R b_L = -Q_R T_L = Q_L T_R ,}
where the last equality is derived from $QT=0$ and the fact that $Q_LT_L=Q_RT_R=0$. Therefore, we have $Q_L(Q_R b_L - T_R)=0$ and there should exist a conformal weight two operator $\Psi$ such that
\eqn\UTY{Q_R b_L - T_R=Q_L\Psi .}
Similarly, starting from $Q_R b_R = T_R$ one obtains
\eqn\UTF{Q_L b_R - T_L=Q_R\widetilde\Psi .}
Adding \UTY\ and \UTF\ one obtains
\eqn\fin{Q_R b_L+Q_L b_R=T+Q_L\Psi+Q_R\widetilde\Psi,}
and choosing $\Psi=-b_L$ and $\widetilde\Psi=-b_R$ one gets the equation 
\eqn\alm{Q_R b_L+Q_L b_R=0.}
Because $Q_L$ is linear in $\l_L$ and
$Q_R$ is linear in $\l_R$, each term in \alm\ has to vanish obtaining \extra. What remains to prove is the consistency between $\pb b_L$ and the BRST transformations. On general grounds it is expected that $\pb b_L$ is $Q_R{\cal O}$ for some $\cal O$ operator. As it was discussed in \BerkovitsZZ, 
$$\pb b_L=\oint_{\Gb} T_R (b_L)=Q_R\left( \oint_{\Gb} b_R(b_L) \right) $$
where $\Gb$ is contour in the ${\bar z}$ plane and $A(B)$ is the OPE of $A$ with $B$. This was proved for the heterotic string case in \FleuryBSD\ and it is expected to be true in the type II case.

\bigskip
\bigskip
\bigskip 
\noindent
{\bf Acknowledgements:} I would like to thank Nathan Berkovits and Brenno Vallilo for valuable comments.

\listrefs

\end